\documentclass[twocolumn,showpacs,preprintnumbers,amsmath,amssymb]{revtex4}
\usepackage[dvips]{graphicx}
\usepackage{float}
\usepackage{pdfpages}
\usepackage{natbib}
\graphicspath{{figures/}}
\usepackage{subfigure}

\begin{document}
\title{Spin-orbit-induced resonances and threshold anomalies in a reduced dimension Fermi gas}
\author{Su-Ju Wang}
\email{wang552@purdue.edu}
\author{Chris H. Greene}
\email{chgreene@purdue.edu}
\affiliation{Department of Physics and Astronomy, Purdue University, West Lafayette, IN
47907, USA}
\date{\today}

\begin{abstract}
We calculate the reflection and transmission probabilities in a one-dimensional Fermi gas with an equal mixing of the Rashba and Dresselhaus spin-orbit coupling (RD-SOC) produced by an external Raman laser field. These probabilities are computed over multiple relevant energy ranges within the pseudo-potential approximation. Strong scattering resonances are found whenever the incident energy approaches either a scattering threshold or a quasi-bound state attached to one of the energetically closed higher dispersion branches. A striking difference is demonstrated between two very different regimes set by the Raman laser intensity, namely between scattering for the single-minimum dispersion versus the double-minimum dispersion at the lowest threshold. The presence of RD-SOC together with the Raman field fundamentally changes the scattering behavior and enables the realization of very different one-dimensional theoretical models in a single experimental setup when combined with a confinement-induced resonance. 

\end{abstract}


\maketitle

Scattering constitutes the fundamental process used to probe countless physical systems, ranging from ultracold dilute gases to energetic quark-gluon plasmas. Rich, intriguing phenomena are found already in one-dimensional quantum scattering processes. For  instance, a quantum particle can tunnel through a double-barrier structure as if no potential exists when the particle energy (even when classically forbidden inside the barrier regions) is resonant with a quasi-bound state supported by the potential. In ultracold atomic systems, a scattering resonance coupling relative motion of two atoms in one or more open channels and bound molecular states in closed channel(s) forms the basis of tunable Fano-Feshbach resonances~\cite{FR}, now extensively used to tune the scattering lengths and enabling many studies in the unitary regime~\cite{Unitary}.

Interactions between two particles can be significantly modified by their external conditions. A recent experiment \cite{Williams} on ultracold atomic collisions in an $^{87}$Rb condensate in the presence of a two-photon Raman field has observed non-spherical scattering halos at very low temperature, where normally only spherically symmetric $s$-wave scattering is expected. The existence of effective higher-partial waves was attributed to the effects of Raman laser dressing. In short, the presence of laser fields modifies collisions between two dressed atoms, and creates effective higher partial wave scattering. Similar experimental setups were used to create spin-orbit coupled Bose-Einstein condensates (BECs) \cite{SOCBEC} and degenerate Fermi gases (DFGs) \cite{SOCFG}\cite{SOCFG2} with an equal mixing of the Rashba and the Dresselhaus spin-orbit coupling. Although SOC appears as a single-particle term in the Hamiltonian, the non-trivial coupling between the internal (spin) with the external (linear momentum) alters the dispersion relation in a fundamental way and thereby changes the intrinsic two-body scattering process.

Of particular interest is the binary collision regime occurring when the scattering energy approaches two crossing points on the positive and negative sides of the lowest SO band. This paper describes the multi-channel scattering resonances in 1D spin-orbit coupled systems controlled by Raman laser fields of moderate strength. In a strong Raman field, the energy band will experience a double-minimum (DM) to single-minimum (SM) transition \cite{FTFD}. These two regimes of physics are found to exhibit qualitatively different 1D models when explored near a confinement-induced resonance \cite{OlshaniiCIR} \cite{SOCCIR}.

\begin{figure}[b!]
\centering
\subfigure{
\includegraphics[width=1.91in]{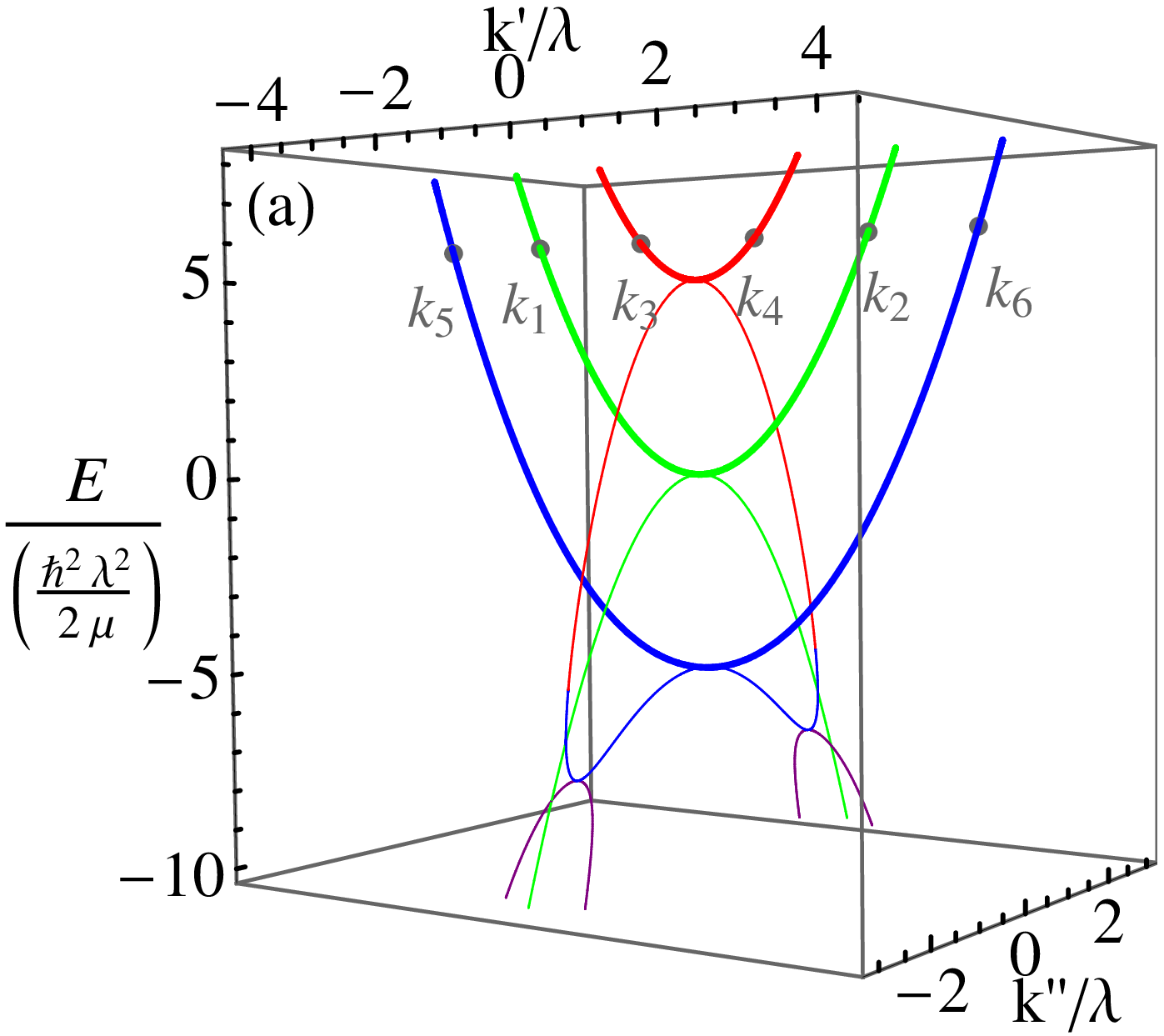}
\label{fig_EkSMA}}
\hspace{-0.24cm}
\subfigure{
\includegraphics[width=0.19\textwidth]{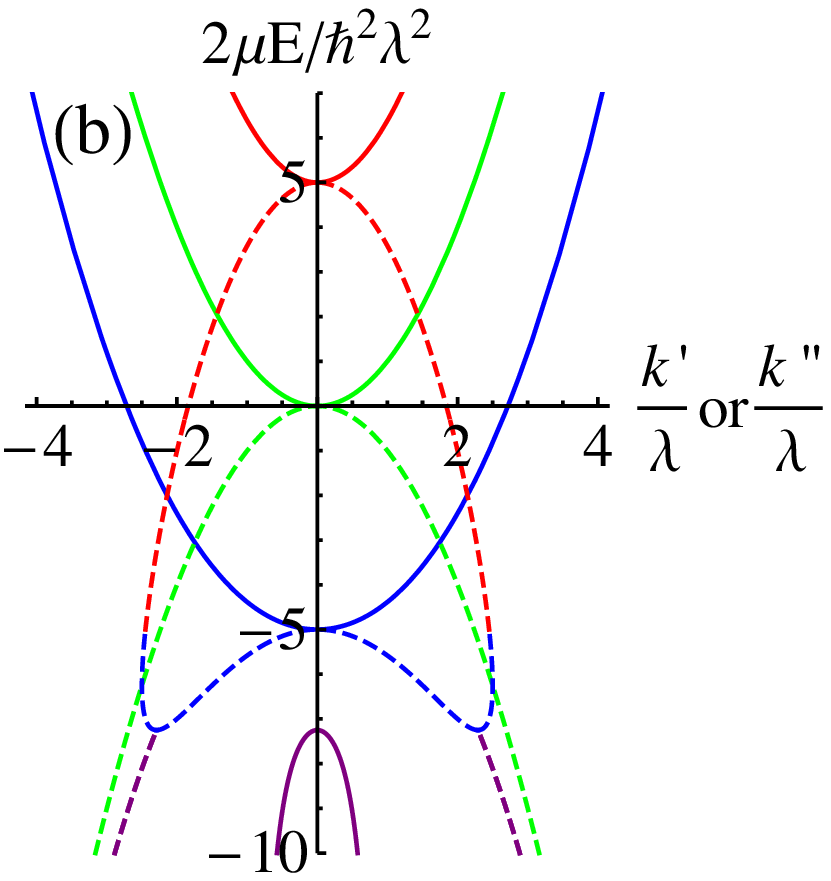}
\label{fig_EkSMB}}
\caption{(a) The total spectrum of states (the relative energy, $E$, versus the wave vector, $k$, where $k'$ is the real part and $k''$ is the imaginary part of $k$) in the single-minimum regime of the center-of-mass frame, i.e. obtained by setting $K=0$. The parameters used are $\hbar\Omega=5\hbar^2\lambda^2/(2\mu)$ and $\hbar\delta=0 \hbar^2\lambda^2/(2\mu)$. The thick curves represent the real wave vector solutions in different dispersion branches shown in thick red, thick green and thick blue. When energy goes below the scattering thresholds, application of the analytical continuation gives us the thin curves, which stand for either the purely imaginary wave vectors (thin red, thin green, and thin blue) or the complex ones (thin purple). (b) A companion to plot (a). The solid (dashed) lines represent the real (imaginary) part of the wave vector, $k$.}
\label{fig_EkSM}
\end{figure}

\begin{figure}[b!]
\centering
\subfigure{
\includegraphics[width=1.91in]{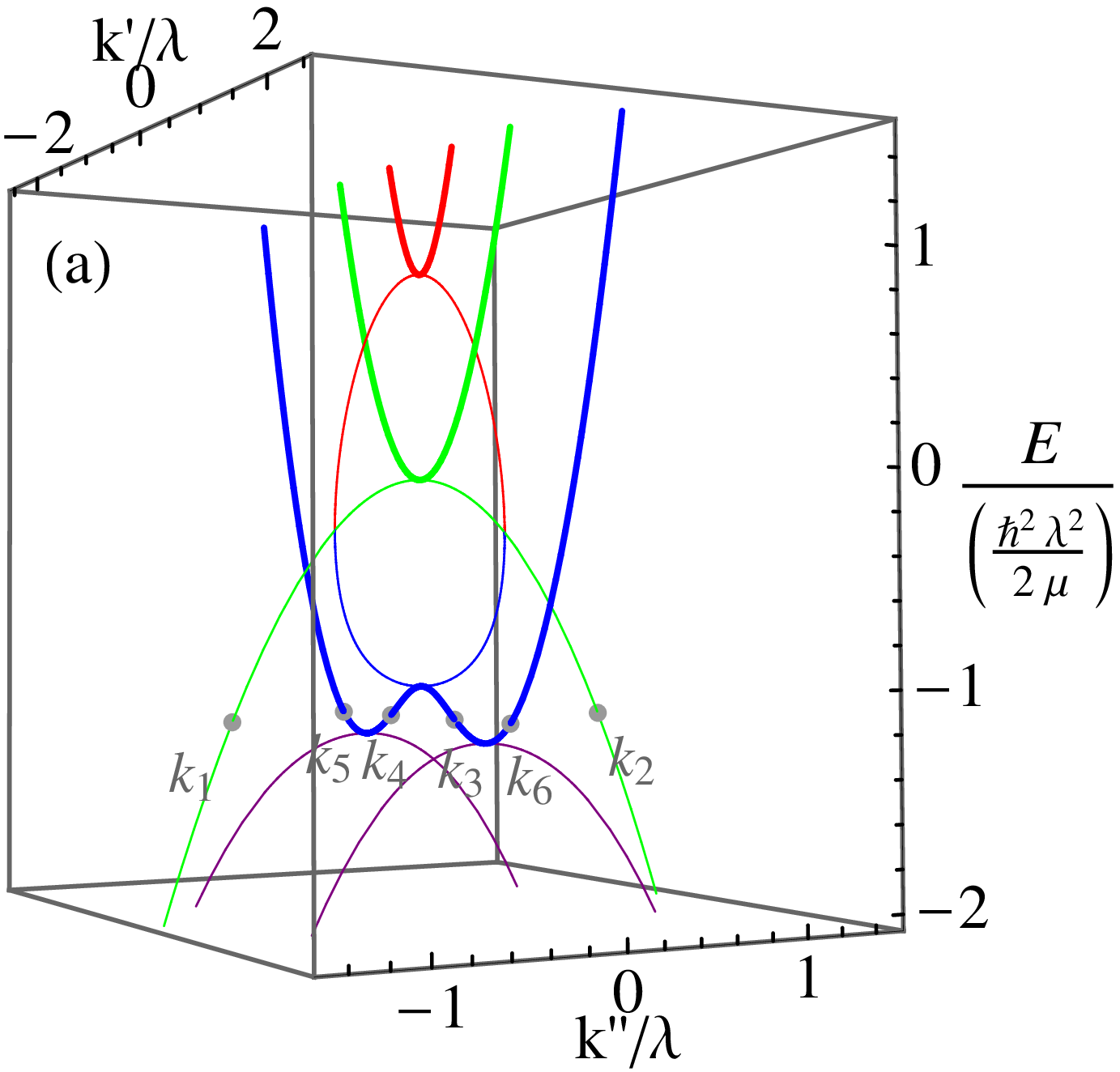}
\label{fig_EkDMA}}
\hspace{-0.24cm}
\subfigure{
\includegraphics[width=0.19\textwidth]{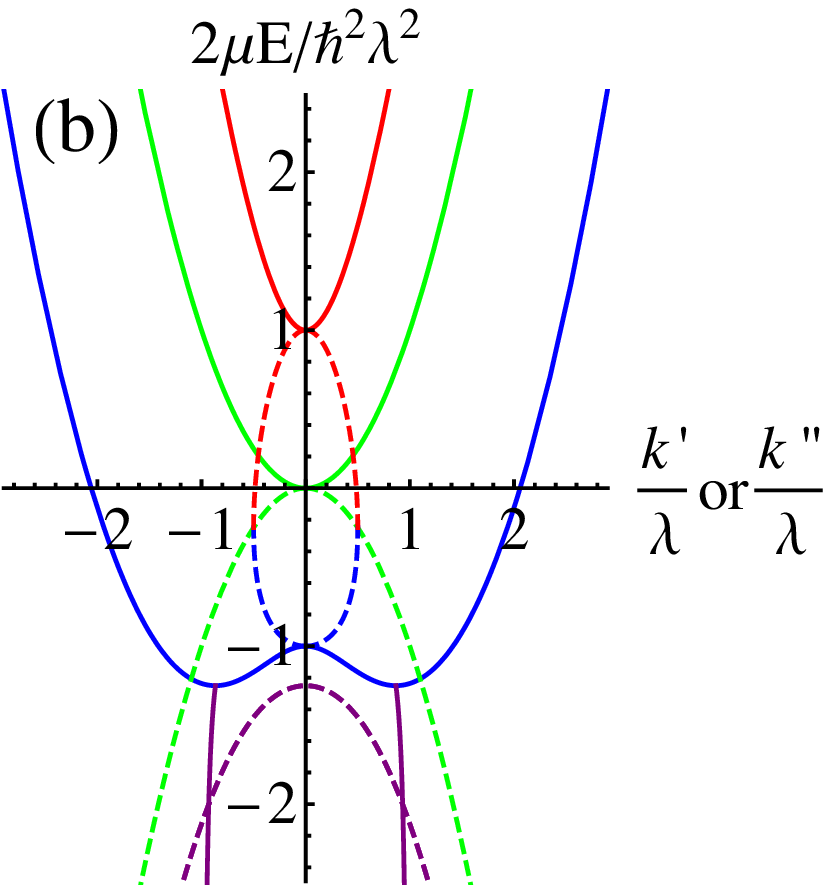}
\label{fig_EkDMB}}
\caption{(a) The total spectrum of states in the double-minimum regime in the center-of-mass frame. Parameters used are $\hbar\Omega=1\hbar^2\lambda^2/(2\mu)$ and $\hbar\delta=0\hbar^2 \lambda^2/(2\mu)$. The thick curves represent real wave vector solutions (thick red, thick green and thick blue). Besides the thin green curve, which is identical to the one in Fig.~\ref{fig_EkSMA}, the purely imaginary solutions (thin red and thin blue) now live between the real solutions. The complex solutions are represented by the thin purple lines. (b) A companion to figure (a). The solid (dashed) lines represent the real (imaginary) part of the wave vector, $k$.}
\label{figEkDM}
\end{figure}

Consider a binary collision as it arises in the experimental protocol of the Zhang group \cite{SOCFG}. The two-body Hamiltonian is
\begin{align}
\label{Eq1}
\nonumber
H_{\text{1D}}=&\frac{\hbar^2k^2_1}{2m}+\frac{\hbar^2\lambda}{m}{k_1} {{\sigma}_{1x}}+ \frac{\hbar\Omega}{2}\sigma_{1z}+\frac{\hbar\delta}{2}\sigma_{1x}+\\
&\frac{\hbar^2{k}^2_2}{2m}+\frac{\hbar^2\lambda}{m}{k}_2{{\sigma}_{2x}}+\frac{\hbar\Omega}{2}\sigma_{2z}+\frac{\hbar\delta}{2}\sigma_{2x}+V(x),
\end{align}
where $\sigma_{i=x,y,z}$ are Pauli matrices for spin-1/2 particles, $\lambda$ is the SOC strength, $\Omega$ is the Raman coupling strength, $\delta$ is the two-photon detuning, and $V(x)$ is the two-body interaction. Because the 1D Rashba-Dresselhaus spin-orbit coupling can be gauged away with a unitary transformation, the RD-SOC term simply causes a constant energy shift for different spin states in the absence of the Raman field and thus does not grant an interesting result. Therefore, the effective magnetic field from the Raman coupling, which is perpendicular to the SOC field, is crucial in our discussions below because it opens a gap between the energy bands, visible in Fig.~\ref{fig_EkSMA} and \ref{fig_EkDMA}. This is very different from the Rashba SOC in 2D or the Weyl SOC in 3D, where the non-abelian nature of the vector potentials alone makes their effects nontrivial.  

After defining the relative momentum and the total momentum as $k=(k_2-k_1)/2$ and $K=k_1+k_2$, we can recast the Hamiltonian in Eq.~(\ref{Eq1}) as:
\begin{align}
\nonumber
H_{\text{1D}}&=\frac{\hbar^2 K^2}{2M} \hat{I} +\frac{\hbar^2 k^2}{2\mu} \hat{I}+ V(x)\\
&
+\left(
\begin{array}{cccc}
0 & \frac{\hbar^2 \sqrt{2} k \lambda}{m}  &   -\frac{\hbar^2 \sqrt{2}k \lambda}{m} & 0\\
  \frac{\hbar^2  \sqrt{2} k \lambda}{m} &   \hbar \Omega&    0  &  \frac{\hbar^2 K\lambda}{\sqrt{2}m}+\frac{\hbar\delta}{\sqrt{2}} \\
  -\frac{\hbar^2  \sqrt{2} k \lambda}{m} &  0  & -\hbar\Omega &  \frac{\hbar^2 K\lambda}{\sqrt{2}m}+\frac{\hbar\delta}{\sqrt{2}}\\
   0 &   \frac{\hbar^2 K\lambda}{\sqrt{2}m} +\frac{\hbar\delta}{\sqrt{2}} &   \frac{\hbar^2 K\lambda}{\sqrt{2}m} +\frac{\hbar\delta}{\sqrt{2}} & 0
\end{array}
\right),
\end{align}
where the matrix is written in the singlet and triplet basis: \{$|S\rangle=(|\uparrow \downarrow \rangle-|\downarrow \uparrow \rangle)/\sqrt{2}, |T_1\rangle=|\uparrow \uparrow \rangle, |T_2\rangle=|\downarrow \downarrow \rangle$, $|T_3\rangle=(|\uparrow \downarrow \rangle+|\downarrow \uparrow \rangle)/\sqrt{2}$\}. These vectors form a complete basis for the Hilbert space of two spin-1/2 particles.
Morover, if we move into the center of mass frame of the two colliding atoms, the triplet channel, $|T_3\rangle$, is decoupled from the other spin channels in the case of zero detuning. Therefore, the dimension of the spin Hilbert space is reduced into three. The corresponding eigenstates and eigenvalues from top to bottom along the energy axis are 

\begin{align}
\label{Es1}
&\Psi_u=\frac{1}{2\sqrt{\hbar^2k^2+\big(\frac{m\Omega}{2\lambda}\big)^2}}
\begin{pmatrix}
    -\sqrt{2} \hbar k \\
     -\sqrt{\hbar^2k^2+\big(\frac{m \Omega}{2\lambda}\big)^2}-\frac{m\Omega}{2\lambda} \\
     \sqrt{\hbar^2k^2+\big(\frac{m\Omega}{2\lambda}\big)^2}-\frac{m\Omega}{2\lambda} \\
\end{pmatrix} e^{ikx}\\
\label{Es2}
&\Psi_m=\frac{1}{\sqrt{2\big[\hbar^2 k^2+\big(\frac{m\Omega}{2\lambda}\big)^2\big]}}
\begin{pmatrix}
    -\frac{m\Omega}{\sqrt{2}\lambda} \\
    \hbar k \\
   \hbar k \\
\end{pmatrix} e^{ikx}\\
\label{Es3}
&\Psi_b=\frac{1}{2\sqrt{\hbar^2k^2+\big(\frac{m\Omega}{2\lambda}\big)^2}}
\begin{pmatrix}
    \sqrt{2} \hbar k \\
     -\sqrt{\hbar^2k^2+\big(\frac{m \Omega}{2\lambda}\big)^2}+\frac{m\Omega}{2\lambda} \\
     \sqrt{\hbar^2k^2+\big(\frac{m\Omega}{2\lambda}\big)^2}+\frac{m\Omega}{2\lambda} \\
\end{pmatrix} e^{ikx}.
\end{align}

\begin{align}
&E_\text{u}=\hbar^2(k^2+\sqrt{4k^2\lambda^2+m^2\Omega^2/\hbar^2})/m,\\
&E_\text{m}=\hbar^2k^2/m,\\
&E_\text{b}=\hbar^2(k^2-\sqrt{4k^2\lambda^2+m^2\Omega^2/\hbar^2})/m.
\end{align}
In the following, we denote the different curves in these computed dispersion relations as "states" or "branches".

The channel structure of the multichannel scattering in the presence of RD-SOC and the Raman field is determined by (1) the incoming scattering energy, $E$, of the relative motion and (2) the relative strength between $(\hbar k/m)\lambda$ and $\Omega$. When the Raman coupling strength is stronger than $\Omega_c=2\hbar\lambda^2/m$, the energy bands are in the single-minimum regime. In this regime, for $-\hbar\Omega<E<0$, there are one open channel and two closed channels. For $0<E<\hbar\Omega$, there are two open channels and one closed channel. For $E>\hbar\Omega$, all channels are open. The channel structure becomes more complex when the Raman strength is weaker than $\Omega_c$ (or stated alternatively, when the RD-SOC strength is significant). In this double-minimum regime, the channel structure is the same as in the single-minimum regime for $E>-\hbar\Omega$. However, the double-minimum scattering threshold, $E^{\text{DM}}_\text{t}=\frac{-\hbar^2\lambda^2}{m}-\frac{m\Omega^2}{4\lambda^2}$, moves below the single minimum threshold, $E^{\text{SM}}_\text{t}=-\hbar\Omega$. Therefore, when $E^\text{DM}_\text{t}<E<-\hbar\Omega$, the double-minimum structure increases the number of open channels into two. The extra open channel indeed comes from the nonexistence of solution in the upper band. The combined fourth-order equation of the most upper and lowest band always gives four algebraic solutions at any given real energy. The description depends on how these four solutions are distributed among these two bands. In a more general description, when the branch point, $E_{\text{br}}=-m\Omega^2/(4\lambda^2)$, which appears due to the square root in the non-quadratic energy branches, is located at a higher energy than the lowest threshold, all the four solutions are associated with the lowest branch. Fig.~\ref{fig_EkSMA} and Fig.~\ref{fig_EkDMA} show all possible states living in different branches in the SM and DM regimes respectively. For a fixed energy, there are six states available, which are labeled by $k_i$ for $i=1,2, ...,6$. The colors are carefully drawn to reflect the origin of branches for each state. For instance, when $E_\text{t}^{\text{DM}}<E<E_\text{br}$, the two states in thin blue come from the analytical continuation of the thick blue branch in Fig.~\ref{fig_EkDMA}.

%

In 1D low-energy collisions, the binary interaction can be well approximated by a contact potential with an effective coupling strength,~ $g_{\text{1D}}$. Assuming a 1D $s$-wave Fermi pseudo-potential, $V(x)=g_{\text{1D}}\delta(x) |S\rangle \langle S|$, the equations that determine the scattering amplitudes can be derived analytically by matching the solutions to the asymptotic boundary conditions with the given channel structures along with the condition of the continuity of the wave function and its derivative. This must be modified in the singlet component, of course, because the contact interaction causes a first derivative discontinuity. Assuming the scattering solutions are 
$\Psi_{R}(x)=\sum_{\alpha=2,4,6} c_\alpha \Psi_\alpha(x)$ and $\Psi_{L}(x)=\sum_{\alpha=1,3,5} c_\alpha \Psi_\alpha(x)$, where the states $\Psi_{\alpha=\text{even (odd)}}$ are chosen to be either right(left)-moving states or decaying states at $x\rightarrow+\infty$ ($x\rightarrow-\infty$). The coefficients can then be found by matching to the following conditions,
\begin{align}
&\Psi_L^{(i)}(x=0)=\Psi_R^{(i)}(x=0)\,\,\,\,\,\;\;\; \forall i=1,2,3\\
&\frac{d\Psi_L^{(i)}}{dx}\bigg|_{x=0^{-}}=\frac{d\Psi_R^{(i)}}{dx}\bigg|_{x=0^{+}} \,\,\,\;\;\;\,\,\, \forall {i=2,3}\\
&\frac{d\Psi_R^{(1)}}{dx}\bigg|_{x=0^+}-\frac{d\Psi_L^{(1)}}{dx}\bigg|_{x=0^-}=\frac{mg_{\text{1D}}}{\hbar^2}\Psi_R(x=0),
\end{align}
where the superscript, $i$, labels the component in the spinor wave functions.
Moreover, due to the existence of spin-orbit coupling, all of the non-interacting states carry a singlet component. Although the states in the closed channel become evanescent (i.e. the wave vectors have an imaginary part) when the scattering energy goes below any channel threshold, the interaction influences the scattering of the propagating states through their coupling to the singlet contact potential.   


\begin{figure*}
\hspace*{0.845in}
\includegraphics[width=0.9\textwidth]{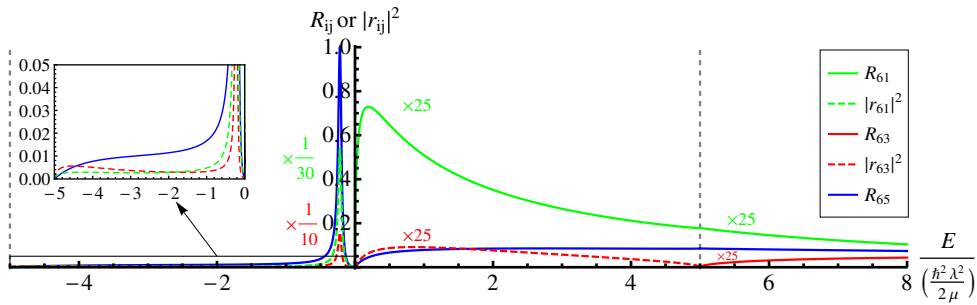}
\caption{The reflection probability/occupation probability as a function of the relative energy, $E$, given the initial state in the blue dispersion branch ($i=6$). The gray dashed vertical lines label the scattering thresholds. When energy goes below a scattering threshold, one branch becomes a closed channel. Therefore, the state becomes evanescent and the amplitude modulus is interpreted as a measure of the occupation probability in the closed channel, which is represented by the dashed curve. Some curves have been rescaled to make them more clearly visible. The multiplication factors are labeled above those curves. The parameters for these calculations are $\hbar\Omega=5\hbar^2\lambda^2/2\mu$, $\delta=0$ and $g_{\text{1D}}=-\hbar\lambda/m$.}
\label{figSMRic6} 
\end{figure*}

\begin{figure}[b!]
\begin{center}
\includegraphics[width=3.3in]{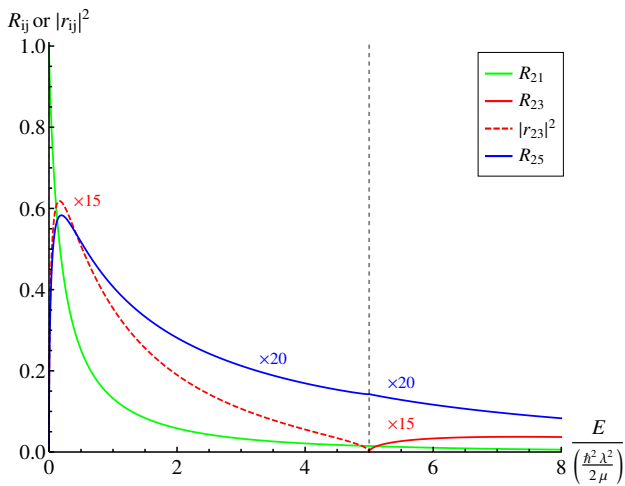}
\caption{The reflection probability/occupation probability is plotted as a function of the relative energy, $E$, given the initial state in the green dispersion branch ($i=2$). The gray dashed vertical lines label the scattering thresholds. When energy decreases below a scattering threshold, one branch becomes energetically closed. Therefore, the state becomes evanescent and the amplitude modulus is interpreted as the occupation probability in the closed channel, which is represented by the dashed curve. Some curves are scaled to be clearly visible. The multiplication factors are labeled above those curves. The parameters used here are $\hbar\Omega=5\hbar^2\lambda^2/2\mu$, $\delta=0$, and $g_{\text{1D}}=-\hbar\lambda/m$.}
\label{figSMRic2} 
\end{center}
\end{figure}

\begin{figure}[b!]
\begin{center}
\includegraphics[width=3.22in]{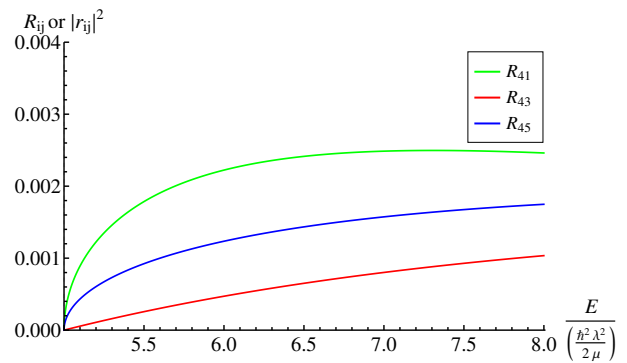}
\caption{The reflection probability is shown as a function of the relative energy, $E$, for an initial state indicated by the green dispersion branch ($i=4$). The parameters for this example are $\hbar\Omega=5\hbar^2\lambda^2/2\mu$, $\delta=0$, and $g_{\text{1D}}=-\hbar\lambda/m$.}
\label{figSMRic4}  
\end{center}
\end{figure}

\begin{figure*}[htp!]
\hspace*{0.845in}
\includegraphics[width=0.9\textwidth]{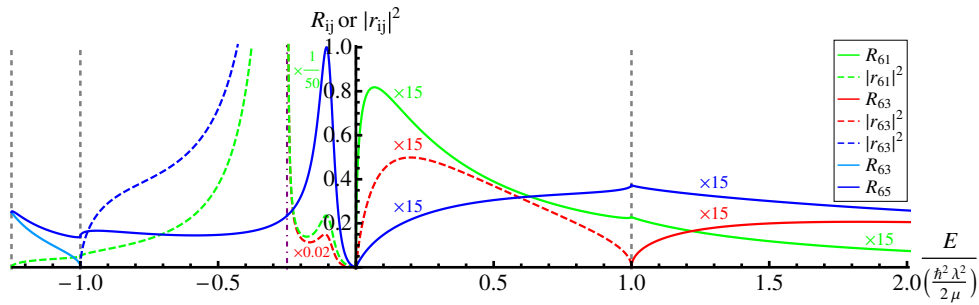}
\caption{The reflection probability/occupation probability as a function of the relative energy, $E$, given the initial state in the blue dispersion branch ($i=6$). The occupation probabilities diverge at the branch point, which is labeled by the dotdashed purple line. The gray dashed vertical lines label the scattering thresholds. When energy goes below a scattering threshold, one branch becomes closed. Therefore, the state becomes evanescent and the amplitude modulus is interpreted as the occupation probability, which is represented by the dashed curve. Some curves are scaled to be clearly visible. The multiplication factors are labeled near those curves. The parameters used here are $\hbar\Omega=1\hbar^2\lambda^2/2\mu$, $\delta=0$, and $g_{\text{1D}}=-\hbar\lambda/m$.}
\label{figDMRic6} 
\end{figure*}

The probability fluxes for the three stationary state components are 
\begin{align}
&v_\text{u}=\frac{2\hbar k}{m}+\frac{4\hbar^2\lambda^2 k/m}{\sqrt{m^2\Omega^2+4\hbar^2\lambda^2 k^2}},\\
&v_\text{m}=\frac{2\hbar k}{m},\\
&v_\text{b}=\frac{2\hbar k}{m}-\frac{4\hbar^2\lambda^2 k/m}{\sqrt{m^2\Omega^2+4\hbar^2\lambda^2 k^2}}.
\end{align}
Under the assumption that the initial state is always one of the non-interacting eigenstates with a positive flux, we can recast the scattering information into reflection probabilities and transmission probabilities characteristic for 1D scattering. For multi-channel scattering, the reflection probability, $R_{if}$, is defined as the ratio of the reflected flux in the outgoing channel, $f$, to the incoming flux in channel, $i$. The transmission probability, $T_{if}$, is found by evaluating the transmitted flux in outgoing channel, $f$, divided by the incoming flux in channel, $i$. The flux conservation guarantees $\sum_f(R_{if}+T_{if})=1$ for a given incoming channel, $i$. The reflection probabilities are plotted versus the relative scattering energy in the single-minimum and double-minimum regimes respectively in Fig.~\ref{figSMRic6}-\ref{figSMRic4} and Fig.~\ref{figDMRic6}.  

When the incident energy increases all the way up to the highest scattering threshold, $E=\hbar\Omega$, the state in the highest branch has a vanishing wave vector (i.e. $k_4=0$) and a zero singlet component. If this state constitutes the incoming state, then the system can be viewed as a non-interacting system since the effective couplings among all scattering states vanish. Therefore, the incoming atoms simply transmit freely through each other and the transmission is 100\% as in the red curve of Fig.~\ref{figSMRic4} at $E=5\hbar^2\lambda^2/2\mu$. In $^{40}$K experiments \cite{SOCFG}, the energy unit, ${\hbar^2\lambda^2/2\mu}$, is about $0.80$$ \mu$K. If the incoming state is any other state with a non-zero wave vector, then the property of the zero probability flux of the state $|k_4\rangle$ causes the vanishing reflection probabilities of channel $|k_4\rangle$ in Fig.~\ref{figSMRic6}, \ref{figSMRic2}, and \ref{figDMRic6} at the highest threshold even under the condition of a non-zero coupling between these scattering states. Above the highest threshold energy, the reflection probability increases with increasing energy because the singlet component of the state $|k_4\rangle$ increases.

Moving down to the next scattering region, where $0<E<\hbar\Omega$, the wave vector of the state $|k_2\rangle$ is again zero at $E=0$. The zero value of the wave vector leads to similar threshold behaviors except for the case with the incoming wave composed of the state $|k_2\rangle$. Since the state with a zero wave vector is a pure singlet state, the middle branch with a normal quadratic dispersion relation scatters as if there is no SOC. Therefore, a total reflection in the incoming channel $|k_2\rangle$ is expected because the incident energy is too weak for particles to pass through, which matches our classical intuition, see the green curve in Fig.~\ref{figSMRic2} at $E=0$. 

\begin{figure}[b!]
\centering
\subfigure{%
\includegraphics[width=0.228\textwidth]{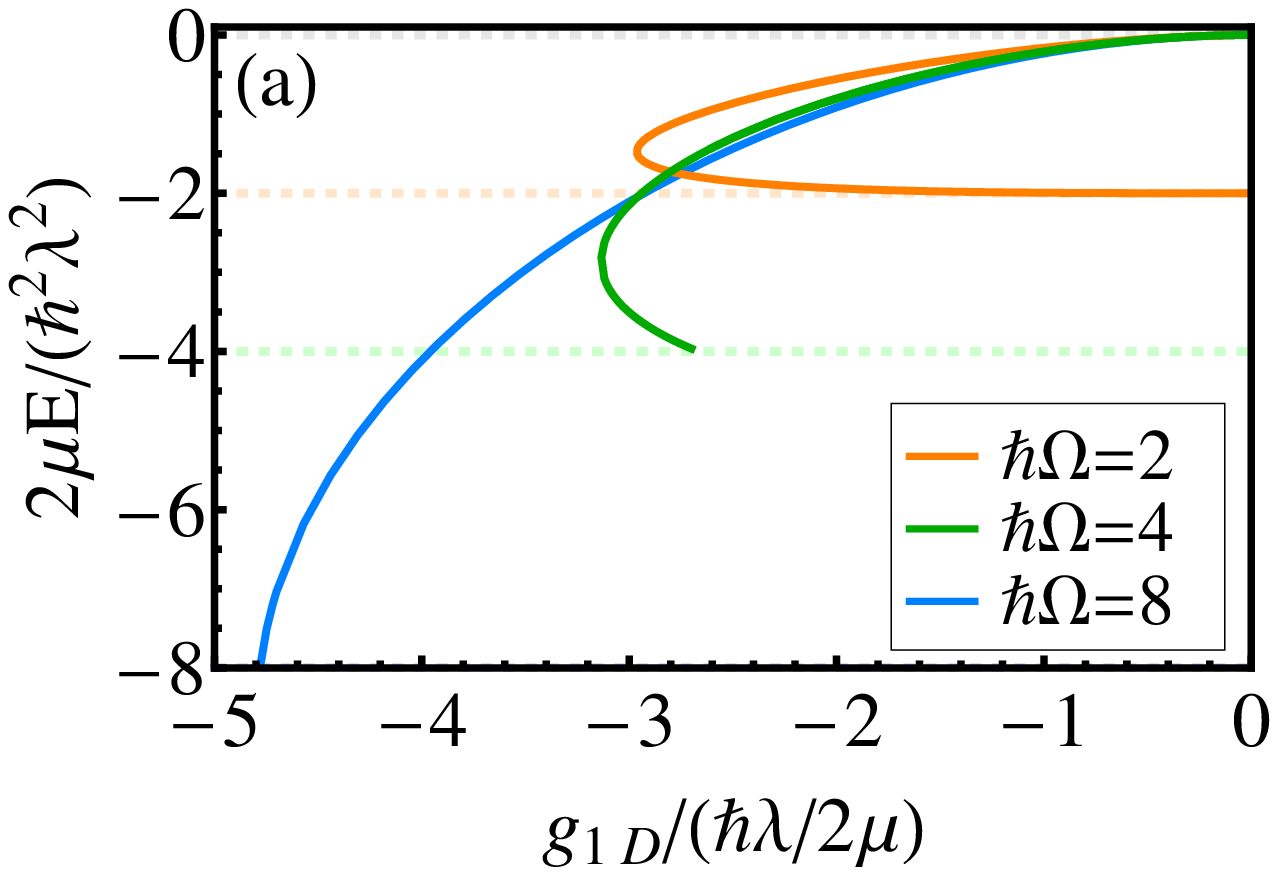}
\label{fig_Res}}
\hspace{-0.4cm}
\subfigure{%
\includegraphics[width=0.228\textwidth]{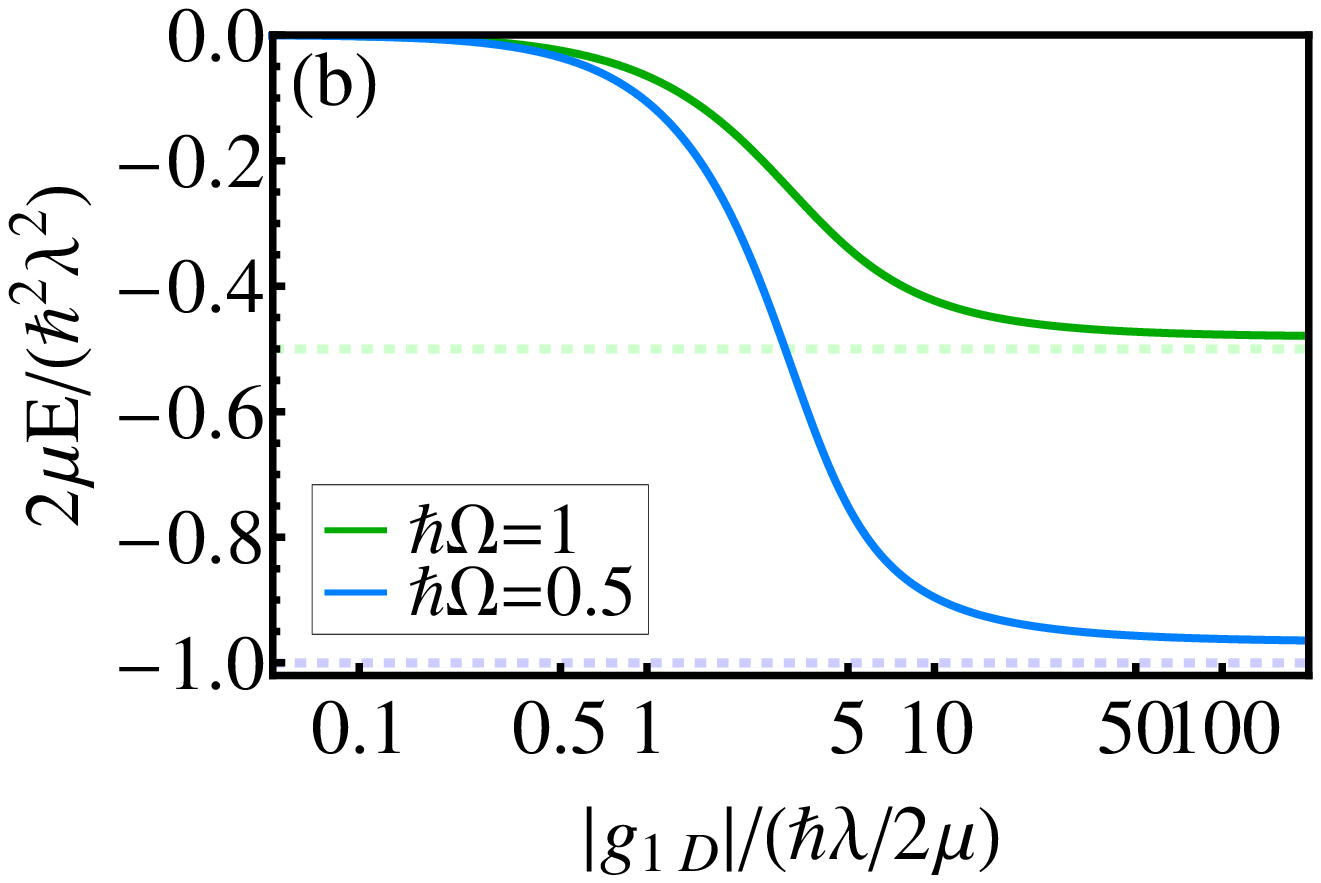}
\label{fig_Res2}}
\hspace*{0.075in}
\subfigure{
\includegraphics[width=0.48\textwidth]{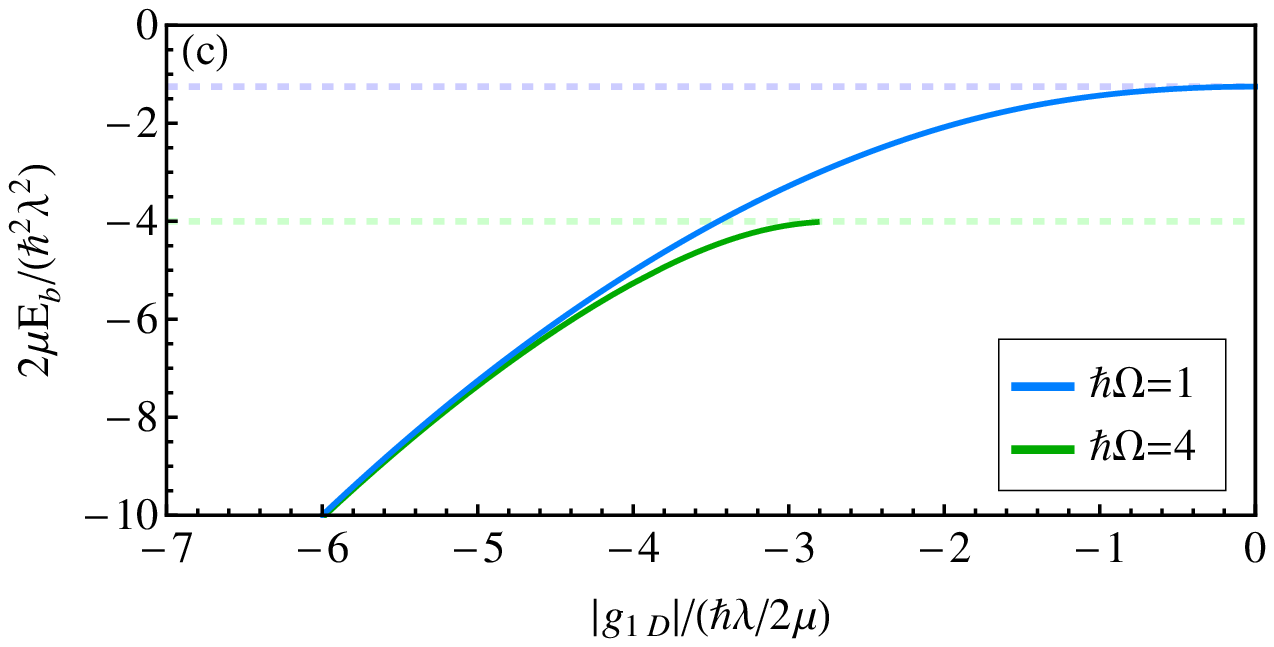}
\label{figBS}}
\caption{Resonance position as a function of $g_{\text{1D}}$ in the energy range $[-\hbar \Omega,0]$, where only one channel is open. (a) The dashed orange, green, and blue line label the lowest scattering threshold for each case. Parameters used are $\hbar\Omega$=2 (orange), 4 (green) and 8 (blue) in the unit of $\hbar^2\lambda^2/2\mu$. (b) Parameters used are $\hbar\Omega=1$ (blue) and $0.5$ (green) in the unit of $\hbar^2\lambda^2/2\mu$. (c) Bound state spectra for $\hbar\Omega=1$ (blue) and $4$ (green) in the unit of $\hbar^2\lambda^2/2\mu$.}
\label{figResboth}
\end{figure}

The more interesting reflection resonances happen when only one open channel exists. Therefore, we extract the resonance position as a function of the interaction strength, $g_\text{1D}$, in the energy range between $[-\hbar\Omega,0]$ in Fig.~\ref{fig_Res} and Fig.~\ref{fig_Res2}. The total reflection shows up as a result of resonance when the scattering energy coincides with the energy of the quasi-bound states supported by the energetically closed bands. Quasi-bound states are possible for the two higher energy bands since their stationary solutions all have singlet components due to the spin-momentum coupling, except at zero momentum. The fact that the evanescent modes have a peaked probability at the same energy position is a strong evidence for the existence of a quasi-bound state \cite{OR}; see the red and green dashed lines around $E=-0.22 \hbar^2\lambda^2/2\mu$ in Fig.~\ref{figSMRic6}. 

\begin{figure}[b!]
\begin{center}
\includegraphics[width=3.2in]{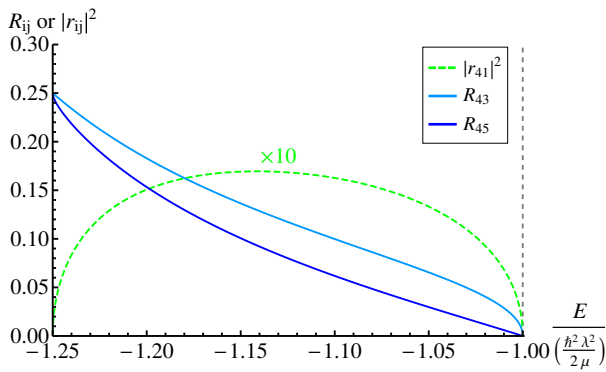}
\caption{The reflection probability in the double-minimum regime when the incoming state has a negative wave vector but a positive flux current ($i=4$). The parameters used in this example are $\hbar\Omega=\hbar^2\lambda^2/2\mu$, $\delta=0$, and $g_{\text{1D}}=-\hbar\lambda/m$. Dashed gray lines are the scattering thresholds.}
\label{figDMRIic4}
\end{center}
\end{figure}

Conventional wisdom suggests that no matter how weak the short-range attraction is, there is always a bound state in 1D quantum physics. This explains the existence of a resonance in Fig.~\ref{fig_Res} for the weakly-interacting region. For instance, it is located between $-2.7\hbar\lambda/m<g_\text{1D}<0$ in the case of $\hbar\Omega=4\hbar^2\lambda^2/2\mu$. When the value of $g_\text{1D}$ is increased, a second resonance peak appears when $-3.1\hbar\lambda/m<g_\text{1D}<-2.7\hbar\lambda/m$ for $\hbar\Omega=4\hbar^2\lambda^2/2\mu$; see the green curve of Fig.~\ref{fig_Res} in the intermediate interaction strength range. The resonance structure reflects the anomalous dispersion of the evanescent modes living between $[-\hbar^2\lambda^2/m-m\Omega^2/(4\lambda^2),-\hbar\Omega]$. When the attraction is strong enough, the scattering process starts to probe the more negative region of the total spectrum, where the double-minimum structure from the evanescent modes contributes to the double peaks in the reflection resonance. The thin blue curve between $[-7.25\hbar^2\lambda^2/(2\mu),-5\hbar^2\lambda^2/(2\mu)]$ in Fig.~\ref{fig_EkSMA} gives a good demonstration of the double-minimum dispersion from the evanescent modes. However, if we increase further the Raman coupling strength, before the attraction is large enough to explore the double-minimum structure of the evanescent branch, the energy of the quasi-bound state has already fallen outside the scattering regime of interest. Therefore, only one peak in the reflection resonance is observed in the blue curve of Fig.~\ref{fig_Res}. 

Switching to the double-minimum regime, see Fig.~\ref{fig_EkDMA}, the evanescent modes from the upper and lower branches now live between energies of the propagating modes which thus leads to very different scattering physics. Because the upper band does not have a solution (or become \textit{transparent}) when the energy goes below $E_{24}=-m\Omega^2/(4\lambda)$, the thin red curve stops at $E_{24}$. Therefore, the occupation probability is replaced by the blue color in Fig.~\ref{figDMRic6} for $E<E_{24}$. In Fig.~\ref{fig_Res2}, we see that there is always only a resonance peak and the peak position is asymptotically approaching $E=-\hbar\Omega$ as $|g_\text{1D}|$ is increased. Due to the completely different topology of the dispersion relation of the evanescent mode, no two resonance peaks at the same $g_\text{1D}$ could be found simultaneously in the DM regime. 

If one inspects the scattering behaviors when the incident energy is at the lowest scattering energy regime in the SM and DM cases, a striking difference is found. For the SM case, there exists a total transmission at the lowest threshold in Fig.~\ref{figSMRic6}; however, for the DM case, there is a partial reflection at $E=E_\text{t}^{\text{DM}}$ in Fig.~\ref{figDMRic6}. The single-minimum regime shares the same threshold behavior as in zero energy due to the decoupling of the incoming channel $|k_6\rangle$ with the other states. In the double-minimum regime, one extra channel is open as $E<-\hbar\Omega$. The extra state carries a flux current, which is in the opposite direction of the wave vector. In Fig.~\ref{figDMRic6}, the scattering between $|k_4\rangle$ and $|k_6\rangle$ cancels out exactly the first derivative discontinuity of the wave function at the threshold, and thus the reflection probability is robust against any change in the interaction strength, $g_{\text{1D}}$. A partial transmission at threshold is explained by the existence of a bound state near the continuum \cite{AThre}. The bound state spectrum has been calculated to support this claim, with evidence shown in Fig.~\ref{figBS}. A bound state is accessible due to the enhancement of the density of states at the lowest threshold energy in the double-minimum regime, and it thus contributes to the partial reflection, which is absent in the normal quadratic dispersion case. In Fig.~\ref{figDMRIic4}, the incoming channel becomes the state $|k_4\rangle$ with a positive flux current but a negative wave vector, analogous to the behavior of light in a metamaterial \cite{meta}. Partial reflection is also observed at the lowest threshold of the energy range of $E_\text{t}^{DM}<E<-\hbar\Omega$ as in Fig.~\ref{figDMRic6}. However, at $E=\hbar\Omega$, the reflection disappears at Fig.~\ref{figDMRIic4} since the channel $|k_4\rangle$ soon becomes closed after crossing that energy.

A quasi-1D system can be realized when the longitudinal kinetic energy is small compared with $2\hbar\omega_\bot$, where $\omega_\bot$ is the transverse trapping frequency. In this limit, the low-energy scattering process has only a total reflection \cite{OlshaniiCIR}\cite{CIRreview} at the scattering threshold as $g_{\text{1D}}\rightarrow \infty$ which leads to the experimental realization of the theoretical 1D Tonks-Girardeau gas \cite{TGgasEXP}\cite{TGgasEXP2} in a strongly repulsive Bose gas. The confinement-induced resonances in the simultaneous presence of the Rashba-Dresselhaus spin-orbit coupling and nonzero Raman coupling are predicted to occur with a less stringent condition of the trapping frequency than the case without RD-SOC upon increasing the magnitude of the Raman coupling strength \cite{SOCCIR}. Our new discoveries of the threshold behaviors in different regimes of Raman coupling in the spin-orbit coupled system extends the capability of using the cold atoms to perform quantum simulations, where very different 1D models could possibly be realized.

This work was supported in part by the National Science Foundation PHY-1607180 and in part by funding from the Purdue University EVPRP. Helpful discussions with Panagiotis Giannakeas and Francis Robicheaux are acknowledged.

\textit{Note}: While preparing the manuscript, the authors noticed that Ref.~\cite{GuanallE} developed a scattering framework which is applicable all the way to the negative energies for an isotropic spin-orbit coupling in 3D. Their treatment should also be applicable to our 1D RD-SOC case and we expect solutions from these two different methods should agree. 

\bibliography{CIRall}
\end{document}